\documentclass[11pt]{article}

\usepackage{amsfonts,amssymb,amsmath, amsthm}
\usepackage{mathrsfs}
\usepackage{pgf}
\usepackage{tikz}
\usetikzlibrary{arrows,automata}
\usepackage[latin1]{inputenc}
\usepackage[all]{xy}
\usepackage{graphicx}
\usepackage{caption}
\usepackage{subcaption}
\usepackage{slashbox}
\usepackage[T1]{fontenc}
\usepackage{relsize}
\usepackage[margin=1in]{geometry}
\usepackage{url}

\usepackage[numbers]{natbib}

\usepackage{color}

\newtheorem{theorem}{Theorem}

\newtheorem{claim}{Claim}
\newtheorem{corollary}{Corollary}
\newtheorem{lemma}{Lemma}
\theoremstyle{definition}
\newtheorem{definition}{Definition}

\theoremstyle{remark}
\newtheorem{remark}{Remark}
\newtheorem{observation}{Observation}
\newtheorem{example}{Example}

\def\blackslug{\hbox{\hskip 1pt \vrule width 8pt height 8pt depth 0pt
\hskip 1pt}}

\def\bqed{\quad\blackslug\lower 8.5pt\null\par}

\def\wqed
{\quad\raisebox{-.3ex}{$\Box$}\lower 8.5pt\null\par}
\long\gdef\boxit#1{\begingroup\vbox{\hrule\hbox{\vrule\kern3pt
      \vbox{\kern3pt#1\kern3pt}\kern3pt\vrule}\hrule}\endgroup}

\newcommand{\cC}{{\cal C}}

\newcommand{\cV}{{\cal V}}

\DeclareMathOperator*{\argmax}{argmax}

\begin{document}



\title{Truthful Multi-unit Procurements with Budgets}



%
%
\author{Hau Chan \quad\quad\quad Jing Chen \\ Department of Computer Science, Stony Brook University\\ \{hauchan, jingchen\}@cs.stonybrook.edu}
%


%

\maketitle
%
%

\begin{abstract}
We study procurement games where each seller supplies multiple units of his item, with a cost per unit known only to him. The buyer can purchase any number of units from each seller, values different combinations of the items differently, and has a budget for his total payment.

For a special class of procurement games, the {\em bounded knapsack} problem, we show that no universally truthful budget-feasible mechanism can approximate the optimal value of the buyer within $\ln n$, where $n$ is the total number of units of all items available. We then construct a polynomial-time mechanism that gives a $4(1+\ln n)$-approximation for procurement games with {\em concave additive valuations}, which include bounded knapsack as a special case. Our mechanism is thus optimal up to a constant factor. Moreover, for the bounded knapsack problem, given the well-known FPTAS, our results imply there is a provable gap between the optimization domain and the mechanism design domain.

Finally, for procurement games with {\em sub-additive valuations}, we construct a universally truthful budget-feasible mechanism that gives an $O(\frac{\log^2 n}{\log \log n})$-approximation in polynomial time with a demand oracle.

\medskip

\end{abstract}

\section{Introduction}\label{sec:intro}
In a procurement game/auction, $m$ sellers compete for providing their items (referred to as products or services in some scenarios) to the buyer.
Each seller $i$ has one item and can provide at most $n_i$ units of it, with a fixed cost $c_i$ per unit which is known only to him.
The buyer may purchase any number of units from each seller.
For example, a local government may buy 50 displays from Dell,
20 laptops from Lenovo, and 30 printers from HP.\footnote{In reality Dell also sells laptops and Lenovo also sells displays. But for the purpose of this paper we consider settings where each seller has one item to supply, but has many units of it.
However, we allow cases where different sellers have the same item, just as one can buy the same laptops from Best Buy and/or Walmart.}
The buyer has a valuation function for possible combinations of the items, 
and a budget $B$ for the total payment he can make. 
We consider universally truthful mechanisms that (approximately) maximize the buyer's value subject to the budget constraint. 


%
Procurement games with budgets have been studied in the framework of budget-feasible mechanisms (see, e.g., \cite{Singer10, Singer11, Ning11, Bei12}).
Yet most studies focus on settings where each seller has only one unit of his item.
Thus there are only two possible allocations for a seller: either his item is taken or it is not.\footnote{In the coverage problem a player has a set of elements,
but still the allocation is bimodal for him: either his whole set is taken or none of the elements is taken.}
When a seller has multiple units and may benefit from selling any number of them,
there are more possibilities for him to deviate into
and it becomes harder to provide incentives for him to be truthful.
To the best of our knowledge, this is the first time where multi-unit budget-feasible mechanisms are systematically studied. 

Multi-unit procurements with budgets can be used to model many interesting problems.
For example, in the classic {\em bounded knapsack} problem the buyer has a value $v_i$ for one unit of item $i$,
and his total value is the sum of his value for each unit he buys.
In job scheduling, the planner may assign multiple jobs to a machine, with different values for different assignments.
As another example, in the Provision-after-Wait problem in healthcare \cite{BCK14}, the government needs to serve $n$ patients at $m$ hospitals. Each patient has his own value for being served at each hospital, and the value of the government is the social welfare.

\subsection{Our main results}

We present our main results in three parts, with most of the proofs provided in the appendix.

\paragraph{An impossibility result.}
Although budget-feasible mechanisms with constant approximation ratios have been constructed
for single-unit procurements 
\cite{Singer10, Ning11},
our first result, formally stated and proved in Section \ref{sec:impossibility},
shows that this is impossible in multi-unit settings, even for the special case of bounded knapsack.

\smallskip

\noindent
{\sc Theorem \ref{thm:impossible}.} (rephrased) {\em No universally truthful, budget-feasible mechanism can do better than a $\ln n$-approximation for bounded knapsack, where $n$ is the total number of units of all items available.}

\smallskip

This theorem applies to all classes of multi-unit procurement games considered in this paper, since they all contain bounded knapsack as a special case.

\paragraph{An optimal mechanism for concave additive valuations.}
A concave additive valuation function is specified by the buyer's marginal values for getting the $j$-th unit of each item $i$, $v_{ij}$, which are  non-increasing in $j$.
The following theorem is formally stated in Section \ref{sec:additive} and proved in Appendix~\ref{app:proof2}.

\smallskip

\noindent
{\sc Theorem \ref{thm:additive}.} (rephrased) {\em There is a polynomial-time mechanism which is a $4(1+\ln n)$-approximation for concave additive valuations.}

\smallskip

Our mechanism is very simple. The central part is a greedy algorithm, which yields a monotone allocation rule. 
However, one needs to be careful about how to compute the payments, and new ideas are needed for
proving budget-feasibility. 

Since bounded knapsack is a special case of concave additive valuations, our mechanism is optimal within a constant factor.
More interestingly, given that bounded knapsack has an FPTAS when there is no strategic considerations,
our results show that there is a gap between the optimization domain
and the mechanism design domain for what one can expect when solving bounded knapsack.

\paragraph{Beyond concave additive valuations.}
We do not know how to use greedy algorithms to construct budget-feasible mechanisms for larger classes of valuations.
The reason is that {\em they may not be monotone:} if a player lowers his cost, he might actually sell fewer units.
This is demonstrated by our example in Section \ref{subsec:subadditive non-monotonicity}. 
Thus we turn to a different approach, {\em random sampling} \cite{Bei12,Goldberg02competitiveauctions,
Dobzinski:2007:TRM:1459737.1459745,journals/corr/abs-1104-2872,
Babaioff:2009:SPW:1496770.1496905,Babaioff:2007:MSP:1283383.1283429}.
The following theorem is formally stated in Section \ref{subsec:sub} and proved in Appendix~\ref{app:proof4}.

\smallskip

\noindent
{\sc Theorem \ref{thm:sub}.} (rephrased) {\em Given a demand oracle, there is a polynomial-time mechanism which is an $O(\frac{\log^2 n}{\log \log n})$-approximation for sub-additive valuations.}

\smallskip

A demand oracle
is a standard assumption for handling sub-additive valuations \cite{Singer11, Bei12, BDO12},
since such a valuation function takes exponentially many numbers to specify.
Notice that for bounded knapsack and concave additive valuations our results are presented using the natural logarithm,
since those are the precise bounds we achieve;
while for sub-additive valuations we present our asymptotic bound under base-2 logarithm, to be consistent with the literature. 


Our mechanism generalizes that of \cite{Bei12}, which gives an $O(\frac{\log n}{\log \log n})$-approximation for  single-unit sub-additive valuations.
Several new issues arise in the multi-unit setting. For example, we must distinguish between an item and a unit of that item,
and in both our mechanism and our analysis we need to be careful about which one to deal with. 
Also, 
we have constructed, as a sub-routine, a mechanism for approximating the optimal {\em single-item outcome}: namely,
an outcome that only takes units from a single seller.
We believe that this mechanism will be a useful building block for budget-feasible mechanisms in multi-unit settings.

\subsection{Related work}


Various procurement games have been studied \cite{Mishra06vickrey-dutchprocurement, PK05, NR01, FPSS02, FSS04}, but without budget considerations.
In particular, frugal mechanisms  \cite{Archer:2007:FPM:1186810.1186813, Cary:2008:ASP:1347082.1347116, Elkind:2004:FPA:982792.982900, Karlin:2005:BVF:1097112.1097495, Talwar:2003:PTF:646517.696327, CEGP10, KSM10} aim at finding mechanisms that minimize the total payment.
As a ``dual'' problem to procurement games, auctions where the buyers have budget constraints have also been studied \cite{AMPP09, GML14}, but the models are very different from ours.
%
%

%
Single-unit budget-feasible mechanisms were introduced by \cite{Singer10}, where the author achieved a constant approximation
for sub-modular valuations.
In \cite{Ning11} the approximation ratio was improved and variants of knapsack problems were studied, but still in single-unit settings. 
In \cite{Singer11} the authors considered single-unit sub-additive valuations and constructed a randomized mechanism that is an $O(\log^2 n)$-approximation and a deterministic mechanism that is an $O(\log^3 n)$-approximation.
We notice that their randomized mechanism can be generalized to multi-unit settings, resulting in an $O(\log^3 n)$-approximation.
In \cite{Bei12} the authors consider both prior-free and Bayesian models. For the former, they provide a constant approximation for XOS valuations and an $O(\frac{\log n}{\log \log n})$-approximation for sub-additive valuations;
and for the latter, they provide a constant approximation for the sub-additive case.
As mentioned we generalize their prior-free mechanism,
but we need to give up a $\log n$ factor in the approximation ratio.
It is nice to see that the framework of budget-feasible mechanism design generalizes to multi-unit settings.

In \cite{NG2013} the author considered settings where each seller has multiple items. 
Although it was discussed why such settings are harder than single-item settings, no explicit upper bound on the approximation ratio was given.
Instead, the focus there was a different benchmark. 
The author provided a constant approximation of his benchmark for sub-modular valuations, but the mechanism does not run in polynomial time.
Also, budget-feasible mechanisms where each seller has one unit of an infinitely divisible item have been considered in \cite{AnariGN14}, under the {\em large-market} assumption: that is, the cost of buying each item completely is much smaller than the budget. The authors constructed a deterministic mechanism which is a $1-1/e$ approximation for additive valuations and which they also prove to be optimal.
In our study we do not impose any assumption about the sellers costs, and the cost of buying all units of an item may or may not exceed the budget.
Moreover, in \cite{Badanidiyuru:2012:LBP:2229012.2229026} the authors studied online procurements and provided a randomized posted-price mechanism that is an $O(\log n)$-approximation for sub-modular valuations under the random ordering assumption.

Finally, knapsack auctions have been studied by \cite{AH06}, where the underlying optimization problem is the knapsack problem, but a seller's private information is the value of his item, instead of the cost. Thus the model is very different from ours and from those studied in the budget-feasibility framework in general.

\subsection{Open problems}
Many questions can be asked about multi-unit procurements with budgets and are worth studying in the future.
Below we mention a few of them.

First, it would be interesting to close the gap between the upper bound in Theorem~\ref{thm:impossible} and the lower bound in Theorem \ref{thm:sub},
even for subclasses such as sub-modular or diminishing-return valuations, as defined in Section \ref{sec:pre}.
A related problem is whether the upper bound 
can be bypassed under other solution concepts.
For example, is there a mechanism with price of anarchy \cite{KP99, RT02} better than $\ln n$?
How about a mechanism with a {\em unique} equilibrium?
Solution concepts that are not equilibrium-based are also worth considering, such as undominated strategies and iterated elimination of dominated strategies.
Another problem is whether a better approximation can be achieved for other benchmarks, such as the one considered in \cite{NG2013}, by truthful mechanisms that run in polynomial time.

Second, online procurements with budget constraints have been studied in both optimization settings \cite{KPV13} and strategic settings \cite{Badanidiyuru:2012:LBP:2229012.2229026}.
But only single-unit scenarios are considered in the latter.
It is natural to ask, what if a seller with multiple units of the same item can show up at different time points, and the buyer needs to decide how many units he wants to buy each time.

Finally, the buyer may have different budgets for different sellers, a seller's cost for one unit of his item may decrease as he sells more,
or the number of units each seller has may  not be publicly known\footnote{In many real-life scenarios the numbers of available units are public information, including procurements of digital products,
procurements of cars, some arms trades, etc.
Here procurement auctions are powerful tools and may result in big differences in prices,
just like in the car market.
However, there are also scenarios where the sellers can hide the numbers of units they have, particularly in a seller's market.
In such cases they may manipulate the supply level, hoping to affect the prices.}.
However, the last two cases are not single-parameter settings and presumably need very different approaches.

%
%
%

\section{Procurement Games}\label{sec:pre}

Now let us define our model. In a procurement game 
there are $m$ {\em sellers} who are the players,  and one {\em buyer}. 
There are $m$ {\em items} and they {\em may or may not be different}. Each player $i$ can provide $n_i$ {\em units} of item $i$,
where each unit is indivisible.
The total number of units of all the items is $n \triangleq \sum_i n_i$.
The {\em true cost} for providing one unit of item $i$ is $c_i\geq 0$, and $c = (c_1,\dots, c_m)$. The value of $c_i$ is player $i$'s private information.
All other information is public. 

An {\em allocation} $A$ is a profile of integers, $A = (a_1,\dots, a_m)$. For each $i\in [m]$, $a_i \in \{0,1,\dots, n_i\}$ and $a_i$ denotes the number of units 
bought from player $i$.
An {\em outcome} $\omega$ is a pair, $\omega = (A, P)$, where $A$ is an allocation and $P$ is the {\em payment profile}:
a profile of non-negative reals with $P_i$ being the payment to player~$i$.
Player $i$'s {\em utility} at $\omega$ is $u_i(\omega) = P_i - a_i c_i$.

The buyer has a {\em valuation function} $V$,  mapping allocations to non-negative reals, such that $V(0, \dots, 0) = 0$.
For allocations $A = (a_1,\dots, a_m)$ and $A' = (a'_1, \dots, a'_m)$ with $a_i\leq a'_i$ for each $i$, $V(A)\leq V(A')$ ---namely, $V$ is {\em monotone}.%
\footnote{Monotonicity is a standard assumption for single-unit budget-feasible mechanisms.}
The buyer has a {\em budget} $B$ and wants to implement an {\em optimal allocation},
$$A^* \in \argmax_{A: \sum_{i\in [m]} c_i a_i \leq B} V(A),$$
while keeping the {\em total payment} within the budget.
An outcome $\omega = (A, P)$ is {\em budget-feasible} if $\sum_{i\in [m]} P_i \leq B$.

\paragraph{The solution concept.}
A deterministic revealing mechanism is {\em dominant-strategy truthful (DST)} if for each player $i$,
announcing $c_i$ is a dominant strategy:
$$u_i(c_i, c'_{-i}) \geq u_i(c'_i, c'_{-i}) \ \forall c'_i, c'_{-i}.$$
A deterministic mechanism is {\em individually rational} if $u_i(c)\geq 0$ for each $i$.
A randomized mechanism is {\em universally truthful} (respectively, {\em individually rational}) if it is a probabilistic distribution over deterministic mechanisms that are DST (respectively, individually rational). 

A deterministic DST mechanism is {\em budget-feasible} if its outcome under $c$ is budget-feasible.
A universally truthful mechanism is {\em budget-feasible (in expectation)} if the expected payment under $c$ is at most $B$.
\begin{definition}
Let $\cC$ be a class of procurement games and $f(n)\geq 0$. 
A universally truthful mechanism is an {\em $f(n)$-approximation} for class $\cC$ if, for any game in $\cC$, the mechanism is individually rational and budget-feasible, and the outcome under the true cost profile $c$ has expected value at least $\frac{V(A^*)}{f(n)}$.
\end{definition}

\begin{remark}
One can trade truthfulness for budget-feasibility: given a universally truthful budget-feasible mechanism,
by paying each player the expected payment he would have received,
we get a mechanism that is {\em truthful in expectation}
and meets the budget constraint {\em with probability 1}.
As implied by Theorem \ref{thm:impossible},
no universally truthful mechanism that meets the budget constraint with probability 1 can do better than a $\ln n$-approximation. Thus there has to be some trade-off. 
\end{remark}

\begin{remark}
We allow different players to have identical items, just like different dealers may carry the same products, with or without the same cost.
But we require the same player's units have the same cost. 
In the future, one may consider cases where one player has units of different items with different costs: that is, a multi-parameter setting instead of single-parameter. 
\end{remark}

Below we define several classes of valuation functions for procurement games.

\paragraph{Concave additive valuations and the bounded knapsack problem.}
An important class of valuation functions are the {\em additive} ones.
For such a function $V$, there exists a value $v_{ik}$ for each item $i$ and each $k\in [n_i]$ such that,
$V(A) = \sum_{i\in [m]}\sum_{k\in [a_i]} v_{ik}$ for any $A = (a_1,\dots, a_m)$.
Indeed, $v_{ik}$ is the marginal value from the $k$-th unit of item $i$ given that the buyer has already gotten $k-1$ units,
no matter how many units he has gotten for other items.
$V$ is {\em concave} if for each $i$, $v_{i1}\geq v_{i2}\geq \cdots \geq v_{in_i}$; namely, the margins for the same item are non-increasing.

A special case of concave additive valuations is the {\em bounded knapsack} problem, one of the most classical problems in computational complexity. Here, all units of an item $i$ have the same value $v_i$: that is, $v_{i1} = v_{i2} = \cdots = v_{i n_i} = v_i$.



\paragraph{Sub-additive valuations.}
A much larger class is the {\em sub-additive} valuations. Here a valuation $V$ is such that, 
for any $A = (a_1,\dots, a_m)$ and $A' = (a'_1, \dots, a'_m)$,
$$V(A\vee A') \leq V(A) + V(A'),$$
where $\vee$ is the item-wise max operation: $A\vee A' = (\max\{a_1, a'_1\}, \dots, \max\{a_m, a'_m\})$.

Notice that the requirement of sub-additivity is imposed only across different players,
and values can change arbitrarily across units of the same player.
Indeed, when $A$ and $A'$ differ at a single player,
sub-additivity {\em does not impose any constraint} on $V(A)$ and $V(A')$,
not even that there are decreasing margins.
Thus this definition is more general than requiring sub-additivity also across units of the same player.
Following the literature, we stick to the more general notion.


Between concave additivity and sub-additivity,
two classes of valuations have been defined, as recalled below.\footnote{The literature of multi-unit procurements has been particularly interested in valuations with some forms of ``non-increasing margins'',
thus has considered classes that contain all concave additive valuations but not necessarily all additive ones.}
To the best of our knowledge, no budget-feasible mechanisms were considered for either of them in multi-unit settings.
\begin{itemize}
\item
{\em Diminishing return}: for any $A$ and $A'$ such that $a_i\leq a'_i$ for each $i$, and for any item~$j$, $V(A+e_j) - V(A) \geq V(A'+e_j) - V(A')$, where $A+e_j$ means adding one extra unit of item $j$ to $A$ unless $a_j= n_j$, in which case $A+e_j = A$.

\item
{\em Sub-modularity}: for any $A$ and $A'$, $V(A\vee A') + V(A\wedge A') \leq V(A) + V(A')$, where $\wedge$ is the item-wise min operation.\footnote{
The item-wise max and min operations when defining sub-additivity and sub-modularity follow directly from the set-union and set-intersection operations
when defining them in general settings, and have been widely adopted in the literature (see, e.g., \cite{CCPV11} and \cite{KPV13}). One may consider alternative definitions where, for example, $\vee$ represents item-wise sum rather than item-wise max. However, we are not aware of existing studies where the alternative definitions are used.}
\end{itemize}

Diminishing return implies sub-modularity,
and both collapse to sub-modularity in single-unit settings.
The reason for diminishing return to be considered separately 
is that
multi-unit sub-modularity is a very weak condition:
when $A$ and $A'$ differ at a single player, it does not impose any constraint,
as sub-additivity. Diminishing return better reflects the idea behind single-unit sub-modularity:
the buyer's value for one extra unit of any item gets smaller as he buys more.

Since the valuation classes defined above are nested:
\begin{eqnarray*}
& & \mbox{bounded knapsack } \subseteq \mbox{concave additivity } \subseteq \mbox{ diminishing return } \\
&\subseteq& \mbox{ sub-modularity } \subseteq \mbox{ sub-additivity},
\end{eqnarray*}
any impossibility result for one class applies to all classes above it, and any positive result for one class applies to all classes below it.
Moreover, since sub-additivity contains additivity, any positive result for the former also applies to the latter.
%

\paragraph{Demand oracle.}
A sub-additive valuation function $V$ may take exponentially many numbers to specify. 
Thus following the studies of single-unit sub-additive valuations \cite{Singer10, Bei12}, we consider a {\em demand oracle}, which takes as input
a set of players $\{1,\dots,m\}$,
a profile of costs $(p_1,\dots, p_m)$ and a profile of numbers of units $(n_1,\dots, n_m)$,\footnote{In single-unit settings a demand oracle takes as input a set of players and the costs.
 For multi-unit settings it is natural to also include the numbers of units.}
and returns, regardless of the budget, an allocation
$$\hat{A} \in \argmax_{A = (a_1,\dots, a_m): a_i\leq n_i \forall i} V(A) - \sum_{i\in [m]}a_i p_i.$$
It is well known that a demand oracle can simulate in polynomial time a value oracle, which returns $V(A)$ given $A$.
Thus we also have access to a value oracle.

\paragraph{Our goal.}
We shall construct universally truthful mechanisms that are individually rational, budget-feasible, and approximate the optimal value of the buyer.
Our mechanisms run in polynomial time for concave additive valuations, and in polynomial time given the demand oracle for sub-additive valuations.



\paragraph{Single-parameter settings with budgets.}
Since the cost $c_i$ is player $i$'s only private information, we are considering single-parameter settings \cite{AT01}. Following Myerson's lemma \cite{doi:10.1287/moor.6.1.58} or the characterization in \cite{AT01}, the only truthful mechanisms are those with a monotone allocation rule and threshold payments. In multi-unit settings, each unit of an item $i$ has its own threshold and the total payment to $i$ will be the sum of the thresholds for all of his units bought by the mechanism.

With a budget constraint, this characterization still holds, but the problem becomes harder: the monotone allocation rule must be such that,
not only (1) it provides good approximation to the optimal value,
but also (2) the unique total payment that it induces must satisfy the budget constraint.
Therefore, similar to single-unit budget-feasible mechanisms, we shall construct monotone allocation rules while keeping an eye on the structure of the threshold payments.
We need to make sure that when the two are combined, both (1) and (2) are satisfied.

\section{Impossibility results for bounded knapsack}\label{sec:impossibility}

The following observation for bounded knapsack is immediate.
\begin{observation} \label{obs1}\hspace{-5pt}{\bf .}
No deterministic DST budget-feasible mechanism can be an $n$-approximation for bounded knapsack.
\end{observation}
\begin{proof}
When $m = 1$, $n_1 = n$, $v_1 = 1$ and $c_1 = B$, a DST mechanism, being an $n$-approximation, must buy 1 unit and pay the player $B$.
When $c_1 = B/n$, the mechanism must still buy 1 unit and pay $B$,
otherwise the player will bid $B$ instead. 
Thus the mechanism's  value is $1$, while the optimal value is $n$. 
\end{proof}

Clearly, buying 1 unit from a player $i \in \argmax_j v_j$ and paying him $B$ is an $n$-approximation. 
For randomized mechanisms we have the following. 
\begin{theorem}\label{thm:impossible}
No universally truthful mechanism can be an $f(n)$-approximation for bounded knapsack with $f(n)<\ln n$.
\end{theorem}
\begin{proof}
%
Consider the case where $m = 1$, $n_1 = n$, and $v_1 = 1$.
For any $b, c\in [0, B]$, let $u_1(b; c)$ be the player's expected utility by bidding $b$ when $c_1 = c$.
For each $k\in [n]$, consider the bid $\frac{B}{k}$: let $P^k$ be the expected payment and, for each $j\in [n]$, let
$p_j^k$ be the probability for the mechanism to buy $j$ units.
When $c_1=\frac{B}{k}$, the optimal value is $k$ and
\begin{equation}\label{equ:1}
\sum_{j\in [n]} p_j^k\cdot j \geq \frac{k}{f(n)} \quad \forall k\in [n],
\end{equation}
as the mechanism is an $f(n)$-approximation.
By universal truthfulness and individual rationality, $u_1(\frac{B}{k}; \frac{B}{k}) \geq u_1(\frac{B}{k-1}; \frac{B}{k}) \ \forall k>1$ and $u_1(B; B) \geq 0$. Namely,

$$P^k - \frac{B}{k}\sum_{j\in [n]} p_j^k\cdot j \geq P^{k-1} - \frac{B}{k} \sum_{j\in [n]} p_j^{k-1}\cdot j \quad \forall k>1, \quad \mbox{and}$$
$$P^1 - B\sum_{j\in [n]} p_j^1\cdot j \geq 0.$$
Summing up these $n$ inequalities, we have
$$\sum_{k\in [n]} P^k - \sum_{k\in [n]} \frac{B}{k}\sum_{j\in [n]} p_j^k\cdot j \geq \sum_{1\leq k<n} P^k - \sum_{1\leq k<n} \frac{B}{k+1} \sum_{j\in [n]} p_j^k \cdot j,$$
which implies
$$P^n \geq \frac{B}{n}\sum_{j\in [n]}p_j^n\cdot j + \sum_{1\leq k<n} \frac{B}{k(k+1)} \sum_{j\in [n]} p_j^k \cdot j.$$
By Equation \ref{equ:1}, we have
$$P^n \geq \frac{B}{f(n)} + \sum_{1\leq k<n} \frac{B}{(k+1)f(n)} = \frac{B}{f(n)}\sum_{k\in [n]}\frac{1}{k} \geq \frac{B \ln n}{f(n)}.$$
By budget-feasibility, $P^n\leq B$. Thus $f(n) \geq \ln n$, implying  
Theorem~\ref{thm:impossible}. 
\end{proof}

\begin{remark}
Notice that as long as the mechanism is truthful {\em in expectation} and individually rational {\em in expectation} (namely, with respect to the players' expected utilities),
the analysis of Theorem \ref{thm:impossible} implies that it cannot do better than a $\ln n$-approximation.
Also notice that the impossibility result does not impose any constraint on the running time of the mechanism. 
\end{remark}

\begin{remark}
When there is a single player, that player has a monopoly and it is not too surprising that no mechanism can do better than a $\ln n$-approximation.
For example, in frugality mechanism design in procurement games, it has been explicitly assumed that there is no monopoly.
However, when monopoly might actually exist, it is interesting to see that there is a tight bound (by Theorems \ref{thm:impossible} and \ref{thm:additive}) on the power of budget-feasible mechanisms in multi-unit settings.
\end{remark}


\section{An optimal mechanism for concave additive valuations}\label{sec:additive}
We construct a polynomial-time universally truthful mechanism $M_{Add}$ that is a $4(1+\ln n)$-approximation for procurement games with concave additive valuations.
Our mechanism is very simple, and the basic idea is a greedy algorithm with proportional cost sharing, as has been used for single-unit settings \cite{Ning11,Singer10}.
However, the key here is to understand the structure of the threshold payments and to show that the mechanism is budget-feasible, which requires ideas not seen before.
Moreover, given our impossibility result, this mechanism is optimal up to a constant factor.
In particular, it achieves the optimal approximation ratio for bounded knapsack. 
The simplicity and the optimality of our mechanism make it attractive to be actually implemented in real-life scenarios.

\paragraph{Notations and Conventions.}
Without loss of generality, we assume $v_{ij}>0$ for each item $i$ and $j\in [n_i]$, since otherwise the mechanism can first remove the units with value 0 from consideration.
Because we shall show that $M_{Add}$ is universally truthful, we describe it only with respect to the truthful bid~$(c_1,\dots, c_n)$.
Also, we describe the allocation rule only, since it uniquely determines the threshold payments. An algorithm for computing the thresholds will be given in the analysis.
Finally, let $i^* \in \argmax_i v_{i1}$ be the player with the highest marginal value, $e_{i^*}$ be the allocation with 1 unit of item $i^*$ and 0 unit of others, and $A_\bot = (0,\dots, 0)$ be the allocation where nothing is bought.
We have the following.


\begin{figure}[tbhp]
\vspace{-10pt}
\centering
\framebox{
\begin{minipage}{\textwidth}{
  \begin{center}
  Mechanism $M_{Add}$ for Concave Additive Valuations
  \end{center}
  \begin{enumerate}
  \item With probability $\frac{1}{2(1+\ln n)}$, go to Step \ref{enu:1}; with probability $\frac{1}{2}$, output $e_{i^*}$ and stop; and with the remaining probability, output $A_\bot$ and stop.
  \item\label{enu:1} For each $i\in [m]$ and $j\in [n_i]$, let the {\em value-rate} $r_{ij} = v_{ij}/c_i$.
  \begin{enumerate}
  \item Order the $n$ pairs $(i,j)$ according to $r_{ij}$ decreasingly, with ties broken lexicographically, first by $i$ and then by $j$.

  For any $\ell\in [n]$, denote by $(i_\ell, j_\ell)$ the $\ell$-th pair in the ordered list.

  \item\label{enu:b}
  Let $k$ be the largest number in $[n]$ satisfying $\frac{c_{i_k}}{v_{i_k j_k}} \leq \frac{B}{\sum_{\ell\leq k}v_{i_\ell j_\ell}}$.

  \item
  Pick up the first $k$ pairs in the list: that is, output allocation $A = (a_1,\dots, a_n)$ where $a_i = |\{\ell: \ell\leq k \mbox{ and }  i_\ell = i\}|$.
  \end{enumerate}
\end{enumerate}
}\end{minipage}
}
\vspace{-10pt}
\end{figure}

\begin{theorem}\label{thm:additive}
Mechanism $M_{Add}$ runs in polynomial time, is universally truthful, and is a $4(1+\ln n)$-approximation for procurement games with concave additive valuations.
\end{theorem}

Theorem \ref{thm:additive} is proved in Appendix \ref{app:proof2}.
Combining Theorems \ref{thm:impossible} and \ref{thm:additive} we immediately have the following.

\begin{corollary}
Mechanism $M_{Add}$ is optimal up to a constant factor among all universally truthful, individually rational, and budget-feasible mechanisms for multi-unit procurement games with concave additive valuations.
\end{corollary}

\begin{remark}
Theorems \ref{thm:impossible} and \ref{thm:additive} show that multi-unit settings are very different from single-unit settings.
In single-unit settings various constant-approximation mechanisms have been constructed, while in multi-unit settings an $O(\log n)$-approximation is the best, and our mechanism provides such an approximation.

Furthermore, for bounded knapsack, without strategic considerations there is an FPTAS, while with strategic considerations the best is a $\ln n$-approximation. Thus we have shown that bound knapsack is a problem for which {\em provably} there is a gap between the optimization domain and the mechanism design domain.

Finally, it would be interesting to see how the constant gap between Theorems \ref{thm:impossible} and \ref{thm:additive} can be closed,
and whether there is a mechanism that meets the budget constraint with probability 1 and achieves an $O(\log n)$-approximation.
\end{remark}

\paragraph{An optimal mechanism for symmetric valuations.}
A closely related class of valuations are the {\em symmetric} ones: there exists $v_1, \dots, v_n$ such that, for any allocation $A$ with $k$ units, $V(A) = \sum_{\ell \leq k} v_\ell$.
In general, symmetric valuations are not concave additive, 
nor are concave additive valuations necessarily symmetric.
But they are  equivalent with a single seller. 
Thus the proof of Theorem \ref{thm:impossible} implies no mechanism can do better than a $\ln n$-approximation for symmetric valuations, as stated in the first part of the theorem below. 
Similar to our analysis of Theorem \ref{thm:additive}, one can verify that the following mechanism is a $4(1+\ln n)$-approximation for symmetric valuations: it is the same as $M_{Add}$ except in Step \ref{enu:1}, where $k$ is set to be the largest number in $[n]$ satisfying $c_{i_k}\leq \frac{B}{k}$. We omit the analysis since it is very similar to that of~$M_{Add}$, and only present the following theorem. 

\begin{theorem}
For symmetric valuations, no universally truthful mechanism can be an $f(n)$-approximation with $f(n)<\ln n$, and there exists a polynomial-time universally truthful mechanism which is a $4(1+\ln n)$-approximation.
\end{theorem}


\section{Truthful mechanisms for sub-additive valuations}\label{sec:subadditive}

\subsection{The non-monotonicity of the greedy algorithm} \label{subsec:subadditive non-monotonicity}
Although the greedy algorithm with proportional cost-sharing played an important role in budget-feasible mechanisms,
we do not know how to use it for multi-unit sub-additive valuations,
since {\em it is not monotone}. 
Indeed, by lowering his cost, a player $i$ will still sell his first unit as in the old allocation.
But once the rank of his first unit changes, all units after that will be re-ranked according to their new marginal value-rates. 
Under the new ordering there is no guarantee whether player~$i$ will sell any of his remaining units.
Below we give an example demonstrating this phenomenon in settings with diminishing returns.

\begin{example}
There are 3 players, $n_1=1$, $n_2 = n_3 = 2$, $c_1 = c_3 = 1$, $c_2 = 1+\epsilon$ for some arbitrarily small $\epsilon>0$, and $B = 3+2\epsilon$.
To highlight the non-monotonicity of the greedy algorithm, 
we work through the algorithm and define the marginal values on the way.
The valuation function will be defined accordingly. 

Given any allocation $A$ and player $i$, denote by $V(i|A)$ the marginal value of item $i$, namely, $V(A+e_i) - V(A)$.
The greedy algorithm works as follows.

\vspace{-4pt}
\begin{itemize}
\item
At the beginning, the allocation is $A_0 = (0, 0, 0)$.


\item
$V(1|A_0) = 10$, $V(2|A_0) = 10+\epsilon$, and $V(3|A_0) = 10-\epsilon$. Item 1 has the largest marginal value-rate, thus $A_1 = (1, 0, 0)$.


\item
$V(1|A_1) = 0$ (item 1 is unavailable now), $V(2|A_1) = 5+5\epsilon$, and $V(3|A_1) = 5-\epsilon$. Item 2 has the largest marginal value-rate, thus $A_2 = (1, 1, 0)$.


\item
$V(1|A_2) = 0$, $V(2|A_2) = 1+\epsilon$, and $V(3|A_2) = 1-\epsilon$. Item 2 has the largest marginal value-rate, thus $A_3 = (1, 2, 0)$.


\item
The budget is used up, the final allocation is $A_3$, and player 2 sells 2 units.
\end{itemize}
\vspace{-4pt}

\noindent
Now let $c'_2 = 1-\epsilon<c_2$. The greedy algorithm works as follows.
\vspace{-4pt}
\begin{itemize}
\item
$A_0 = (0, 0, 0)$.

\item
$V(1|A_0) = 10$, $V(2|A_0) = 10+\epsilon$, and $V(3|A_0) = 10-\epsilon$. Item 2 has the largest marginal value-rate, thus $A'_1 = (0, 1, 0)$.

Notice that player 2 sells his first unit earlier than before.

\item
$V(1|A'_1) = 5+4\epsilon$, $V(2|A'_1)=5-5\epsilon$, and $V(3|A'_1) = 5+5\epsilon$. Item 3 has the largest marginal value-rate, thus $A'_2 = (0, 1, 1)$.

\item
$V(1|A'_2) = 1-2\epsilon$, $V(2|A'_2) = 1-\epsilon$, and $V(3|A'_2) = 1+\epsilon$, thus $A'_3 = (0, 1, 2)$.

\item
The remaining budget is $3\epsilon$, no further unit can be added, and the final allocation is $A'_3$. But player 2 only sells one unit, violating monotonicity.
\end{itemize}
\vspace{-4pt}
%


\noindent
Given the marginal values, the valuation function is defined as follows:
\begin{eqnarray*}
& & V(0, 0, 0) = 0,
 V(1, 0, 0) = 10, V(0, 1, 0) = 10+\epsilon, V(0, 0, 1) = 10-\epsilon, \\
& & V(1, 1, 0) = 15+5\epsilon, V(1, 0, 1) = 15-\epsilon, V(0, 2, 0) = 15-4\epsilon, V(0, 1, 1) = 15+6\epsilon, \\
& & V(0, 0, 2) = 15, \\
& & V(1, 2, 0) = 16+6\epsilon, V(1, 1, 1) = 16+4\epsilon, V(1, 0, 2) = 16, V(0, 2, 1) = 16+5\epsilon, \\
& & V(0, 1, 2) = 16+7\epsilon, \\
& & V(0, 2, 2) = V(1, 2, 1) = V(1, 1, 2) = 16+7\epsilon,
 V(1, 2, 2) = 16+7\epsilon.
\end{eqnarray*}
One can verify that $V$ is consistent with the marginal values and has diminishing returns.
Indeed, for any allocation with $k$ units for $k$ from $0$ to $4$, the marginal value of adding 1 more unit is roughly $10, 5, 1, \epsilon, 0$, and thus diminishing.

\end{example}

Given the non-monotonicity of the greedy algorithm, we turn to another approach for constructing truthful mechanisms, namely, {\em random sampling}.
We  provide our main mechanism in Section \ref{subsec:sub}. In Section \ref{sec:one} we first construct a mechanism that will be used as a subroutine.

\subsection{Approximating the optimal single-item allocation}\label{sec:one}

%

From the analysis of Theorem \ref{thm:impossible}, we notice that part of the hardness in designing mechanisms for multi-unit settings comes from cases where a single player's item contributes a lot to the optimal solution.
In order to obtain a good approximation, we need to identify such a player and buy {\em as many units as possible from him}.
More precisely, given the true cost profile $(c_1,\dots, c_n)$, let
$$i^{**}\in \argmax_i V(\min\{n_i, \lfloor\frac{B}{c_i}\rfloor\} \cdot e_i),$$
where for any $\lambda\in [n_i]$, $\lambda e_i$ is the allocation with $\lambda$ units of item $i$ and 0 unit of others.
Ideally we want to buy $\lambda^{**} \triangleq \min\{n_{i^{**}}, \lfloor\frac{B}{c_{i^{**}}}\rfloor\}$ units from $i^{**}$ and pay him (at most) $B$.
We shall refer to $(i^{**}, \lambda^{**})$ as the {\em optimal single-item allocation}.

Notice that a similar scenario occurs in single-unit settings:
part of the value approximation comes from a single player $i^*$ with the highest marginal value. 
%
The problem is, although the identity of player $i^*$ is publicly known,
both $i^{**}$ and $\lambda^{**}$ depend on the players' true costs and have to be solved from their bids.
Below we construct a universally truthful mechanism, $M_{One}$,
which is budget-feasible and approximates $V(\lambda^{**} e_{i^{**}})$ within a $1+\ln n$ factor.
We have the following theorem, proved in Appendix \ref{app:proof4}.

%
%
%
%
%
%
%
%
%
%
%

\begin{figure}[bthp]
\centering
\framebox{
\begin{minipage}{\textwidth}{
\begin{center}
Mechanism $M_{One}$ for Approximating the Optimal Single-item Allocation
\end{center}

With probability $\frac{1}{1+\ln n}$, do the following.

\begin{enumerate}

\item\label{step:one 1}
Let $v_i = V(\min\{n_i, \lfloor\frac{B}{c_i}\rfloor\} \cdot e_i)$ and order the players according to the $v_i$'s decreasingly, with ties broken lexicographically.

Let $i^{**}$ be the first player in the list and $\lambda^{**} =\min\{n_{i^{**}}, \lfloor\frac{B}{c_{i^{**}}}\rfloor\}$.

\item\label{step:one 2}
Let $k\in [\lambda^{**}]$ be the smallest number such that player $i^{**}$ is still ordered the first with cost $c'_{i^{**}} = \frac{B}{k}$. 

\item\label{step:one 3}
Set $\theta_\ell = \frac{B}{k}$ for each $\ell\leq k$ and $\theta_\ell = \frac{B}{\ell}$ for each $k+1 \leq \ell \leq \lambda^{**}$.

\item\label{step:one 4}
Output allocation $\lambda^{**} e_{i^{**}}$ and pay $\sum_{\ell \leq \lambda^{**}}\theta_\ell$ to player $i^{**}$.
\end{enumerate}

}
\end{minipage}
}
\vspace{-10pt}
\end{figure}

\begin{theorem}\label{lem:one}
Mechanism $M_{One}$ is universally truthful, individually rational, budget-feasible, and is a $(1+\ln n)$-approximation for $V(\lambda^{**} e_{i^{**}})$.
\end{theorem}

Since the impossibility result in Theorem \ref{thm:impossible} applies to settings with a single item, we have the following corollary.
\begin{corollary}
Mechanism $M_{One}$ is optimal for approximating $V(\lambda^{**} e_{i^{**}})$ among all universally truthful, individually rational, and budget-feasible mechanisms.
\end{corollary}

\begin{remark}
As it will become clear from the analysis, $M_{One}$ does not require the valuation to be sub-additive.
The only thing it requires is that,
for each player~$i$, $V(\lambda e_i)$ is non-decreasing in $\lambda$.
Thus it can be used for valuations that are not even monotone, as long as they are monotone across units of the same item. 

Furthermore, given that $(i^{**}, \lambda^{**})$ is the multi-unit counterpart of player $i^*$ in single-unit settings, and given the important role $i^*$ has played in
single-unit budget-feasible mechanisms, we believe mechanism $M_{One}$ will be a useful building block in the design of budget-feasible mechanisms for multi-unit settings.

\end{remark}

\subsection{A truthful mechanism for sub-additive valuations}\label{subsec:sub}

Our mechanism for sub-additive valuations generalizes that of  \cite{Bei12}. 
In particular, the algorithm~$A_{Max}$ and the mechanism $M_{Rand}$ below are respectively variants of their algorithm {\sc SA-alg-max} and mechanism {\sc SA-random-sample}.
Several new issues arise in multi-unit settings.
For example, we must now distinguish between an item and a unit of that item.
In the mechanism and its analysis, we sometimes deal with an item ---thus all of its units at the same time--- and sometimes deal with a single unit.
Also, as discussed in Section \ref{sec:one},
the role of player~$i^*$ with the highest marginal value is replaced by player $i^{**}$, and the way $i^{**}$ contributes to the value approximation has changed a lot ---this is where the extra $\log n$ factor comes.
Indeed, to construct and analyze our mechanism one need good understanding of the problem in multi-unit settings.
Our mechanism $M_{Sub}$ is a uniform distribution between $M_{Rand}$ and the mechanism~$M_{One}$ of Section \ref{sec:one}.
We have the following theorem, proved in Appendix \ref{app:proof4}.
%
%


\begin{figure}[tbhp]
\centering
\framebox{
\begin{minipage}{\textwidth}{
\begin{center}
Algorithm $A_{Max}$
\end{center}

Since this algorithm will be used multiple times with different inputs, we specify the inputs explicitly to avoid confusion.
Given players $1,\dots, m$, numbers of units $n_1, \dots, n_m$, costs $c_1,\dots, c_m$, budget $B$, and a demand oracle for the valuation function $V$, do the following.

\begin{enumerate}
\item
Let $n'_i = \min\{n_i, \lfloor\frac{B}{c_i}\rfloor\}$ for each $i$, $i^{**} = \argmax_i V(n'_i e_i)$, 
$v^* = V(n'_{i^{**}} e_{i^{**}})$, and $\cV = \{v^*, 2v^*, \dots, mv^*\}$.
\item
For $v\in \cV$ from $mv^*$ to $v^*$,
\begin{enumerate}

\item
Set $p_i = \frac{v}{2B}\cdot c_i$ for each player $i$. Query the oracle with $m$ players, number of units $n'_i$ and cost $p_i$ for each $i$, to find

$S = (s_1,\dots, s_m) \in \arg\max_{A = (a_1,\dots, a_m): a_i\leq n'_i \forall i} V(A)-\sum_{i\in [m]} a_i p_i$.

(When there are multiple optimal solutions, the oracle always returns the same one whenever queried with the same instance.)

\item
Set allocation $S_v = A_\bot$. (Recall $A_\bot = (0,\dots, 0)$ represents buying nothing.)

\item
If $V(S) < \frac{v}{2}$, then continue to the next $v$.

\item
Else, order the players according to $s_i c_i$ decreasingly with ties broken lexicographically, and denote them by $i_1,\dots, i_m$.

Let $k$ be the largest number in $[m]$ satisfying $\sum_{\ell\leq k} s_{i_\ell} c_{i_\ell}\leq B$,
and let $S_v$ be $S$ projected on $\{i_1,\dots, i_k\}$:
$S_v = \bigvee_{\ell\leq k} s_{i_\ell} e_{i_\ell}$, namely, $S_v$ consists of taking $s_{i_\ell}$ units of item $i_\ell$ for each $\ell\leq k$, and taking 0 unit of others.

\end{enumerate}

\item
Output $S_{Max} \in \argmax_{v\in \cV} V(S_v)$.

(When there are several choices, the algorithm chooses one arbitrarily, but always outputs the same one when executed multiple times with the same input.)

\end{enumerate}
}
\end{minipage}
}
\end{figure}

\vspace{-10pt}
\begin{figure}[tbhp]
\vspace{-10pt}
\centering
\framebox{
\begin{minipage}{\textwidth}{
\begin{center}
Mechanism $M_{Rand}$
\end{center}

\begin{enumerate}

\item\label{enu:rand-1}
Put each player independently at random with probability $1/2$ into group $T$, and let $T' = [m]\setminus T$.

\item\label{enu:rand-2}
Run $A_{Max}$ with the set of players $T$, number of units $n_i$ and cost $c_i$ for each $i\in T$, budget $B$, and the demand oracle for valuation function $V$.
Let $v$ be the value of the returned allocation.

\item
For $k$ from $1$ to $\sum_{i\in T'} n_i$,

\begin{enumerate}

\item
Run $A_{Max}$ with the set of players $T_k = \{i : i\in T', c_i\leq \frac{B}{k}\}$, number of units $n_i$ and cost $\frac{B}{k}$ for each $i\in T_k$, budget $B$, and the demand oracle for $V$.
Denote the returned allocation by $X = (x_1, \dots, x_m)$, where $x_i = 0$ for each $i\notin T_k$.

\item\label{enu:rand-3b}
If $V(X) \geq \frac{\log \log n}{64 \log n} \cdot v$, then output allocation $X$, pay $x_i \cdot \frac{B}{k}$ to each player $i$, and stop.

\end{enumerate}

\item
Output $A_\bot$ and pay 0 to each player.

\end{enumerate}
}
\end{minipage}
}
\end{figure}


\begin{theorem}\label{thm:sub}
Mechanism $M_{Sub}$ runs in polynomial time, is universally truthful, and is an $O(\frac{(\log n)^2}{\log \log n})$-approximation for procurement games with sub-additive valuations.
\end{theorem}

Since diminishing return, sub-modularity, and additivity all imply sub-additivity, we immediately have the following.

\begin{corollary}\label{col:add}
$M_{Sub}$ is an $O(\frac{(\log n)^2}{\log \log n})$-approximation for procurement games with diminishing returns, those with sub-modular valuations, and those with additive valuations.
\end{corollary}

\begin{remark}
The worst case of the approximation above comes from cases where $V(\lambda^{**} e_{i^{**}})$ (and thus $M_{One}$) is the main contribution to the final value.
Unlike single-unit settings, we need an additional $\log n$ factor because the optimal approximation ratio for $V(\lambda^{**} e_{i^{**}})$ is $O(\log n)$.
For scenarios where the players' costs are very small, in particular, where $n_i c_i\leq B$ for each $i$, the optimal single-item allocation $(i^{**}, \lambda^{**})$ is publicly known, just as the player $i^*$ in single-unit settings.
In such a {\em small-cost} setting, which is very similar to the {\em large-market} setting considered by \cite{AnariGN14} except that the items here are not infinitely divisible,
the subroutine $M_{One}$ in $M_{Sub}$ can be replaced by ``allocating $n_{i^{**}}$ units of item $i^{**}$ and paying him $B$'', and the $\log n$ factor is avoided, resulting in an $O(\frac{\log n}{\log \log n})$-approximation.

A small-cost setting is possible in some markets, but it is not realistic in many others. For example, in the Provision-after-Wait problem in healthcare \cite{BCK14}, it is very unlikely that all patients can be served at the most expensive hospital within the government's budget.
Also, in many procurement games, a seller, as the manufacture of his product, can be considered as having infinite supply,
and the total cost of all units he has will always exceed the buyer's budget.
Thus one need to be careful about where the small-cost condition applies. 

\end{remark}
\section*{Acknowledgements} 
We thank several anonymous reviewers for their comments. The first author is supported by the NSF Graduate Research Fellowship.

\bibliographystyle{plain}

%
%
%
%
%
%

\appendix

\section{Proof of Theorem \ref{thm:additive}}\label{app:proof2}

We break the proof of Theorem \ref{thm:additive} into a sequence of lemmas. 

\begin{lemma}\label{lem:dst}
Mechanism $M_{Add}$ is universally truthful and individually rational.
\end{lemma}
\begin{proof}
This mechanism is a probabilistic distribution over three deterministic sub-mechanisms: the one outputs~$A_\bot$, the one outputs $e_{i^*}$, and the one in Step~\ref{enu:1}. Obviously the first two are monotone.

For the one in Step \ref{enu:1}, notice that for any $k'\leq k$ we have
$$\frac{c_{i_{k'}}}{v_{i_{k'}j_{k'}}} \leq \frac{c_{i_k}}{v_{i_k j_k}} \leq \frac{B}{\sum_{\ell\leq k}v_{i_\ell j_\ell}} \leq \frac{B}{\sum_{\ell\leq k'}v_{i_\ell j_\ell}},$$
and for any $k'\geq k+1$ we have
$$\frac{c_{i_{k'}}}{v_{i_{k'}j_{k'}}} \geq \frac{c_{i_{k+1}}}{v_{i_{k+1}j_{k+1}}} > \frac{B}{\sum_{\ell\leq k+1}v_{i_\ell j_\ell}} \geq \frac{B}{\sum_{\ell\leq k'}v_{i_\ell j_\ell}}.$$
Thus a pair $(i_{k'}, j_{k'})$ is picked up ---namely, the $j_{k'}$-th unit of player $i_{k'}$ is bought--- if and only if
\begin{equation}\label{equ:2}
\frac{c_{i_{k'}}}{v_{i_{k'}j_{k'}}} \leq \frac{B}{\sum_{\ell\leq k'}v_{i_{\ell}j_{\ell}}}.
\end{equation}
For any player $i$ and pair $(i,j)$ picked up by the mechanism, when $c_i$ decreases, $(i, j)$'s rank is smaller than or equal to its previous rank. Since $(i, j)$ satisfied Inequality \ref{equ:2} before,
it continues to  satisfy under the new ordering because the left-hand side becomes smaller and the right-hand side can only become larger. Accordingly, $(i,j)$ will still be picked up under the new ordering. Therefore the number of units of player $i$ bought by the mechanism will never decrease when $i$'s cost decreases. Hence the mechanism is monotone.

Since each sub-mechanism pays the players according to the thresholds, we have that each one of the sub-mechanisms is DST. Thus $M_{Add}$ is universally truthful.

It is easy to see that $M_{Add}$ is individually rational. Indeed, the sub-mechanism outputting $A_\bot$ pays 0 to every player and the sub-mechanism outputting $e_{i^*}$ pays $B$ to player $i^*$ and 0 to others. Thus the players get non-negative utilities in both of them. For the sub-mechanism in Step \ref{enu:1} and for any player $i$, even without giving the explicit formula of the threshold payments, by the monotonicity of the allocation rule one can see that,
any $j$-th unit of player $i$ bought under $i$'s true cost $c_i$ will still be bought under any cost $c_i'<c_i$,
and thus the threshold for buying the $j$-th unit of player $i$ is at least $c_i$. Accordingly, the total payment to $i$ is at least $c_i a_i$, giving player $i$ a non-negative utility.
In sum, $M_{Add}$ is individually rational and Lemma \ref{lem:dst} holds. 
\end{proof}

\begin{lemma}\label{lem:poly}
Mechanism $M_{Add}$ runs in polynomial time.
\end{lemma}

\begin{proof}
The only thing that is not clear from the mechanism's description is how to compute the threshold payments for the sub-mechanism in Step 2.
For each player $i$ and each $j\leq a_i$, letting $\theta_{ij}$ be the threshold for the $j$-th unit of $i$, we compute $\theta_{ij}$ using the following algorithm $A_{Th}$.
\begin{center}
\framebox{
\begin{minipage}{\textwidth}{
\begin{center}
Algorithm $A_{Th}$ for Computing the Threshold $\theta_{ij}$

\begin{enumerate}

\item
Order the $n-n_i$ pairs $(i', j')$ with $i'\neq i$ according to the value-rate $r_{i'j'}$'s decreasingly, with ties broken lexicographically.
Denote by $(i'_\ell, j'_\ell)$ the $\ell$-th pair in the list.

\item
Set $t'_{n-n_i+1} = +\infty$.

\item
For $\alpha$ from $n-n_i$ to 0, compute $t_\alpha = \frac{v_{ij} B}{\sum_{\ell\leq j}v_{i\ell} + \sum_{\ell\leq \alpha}v_{i'_\ell j'_\ell}}$ and $t'_\alpha = \frac{v_{ij} c_{i'_\alpha}}{v_{i'_\alpha j'_\alpha}}$, except for $\alpha=0$, where $t'_0 = 0$.
\begin{enumerate}
\item\label{enu:a}
If $t_\alpha< t'_\alpha$, continue to the next round.

\item\label{enu:3b}
If $t'_\alpha \leq t_\alpha\leq t'_{\alpha+1}$, set $\theta_{ij} = t_\alpha$ and stop.

\item\label{enu:3c}
If $t_\alpha> t'_{\alpha+1}$, set $\theta_{ij} = t'_{\alpha+1}$ and stop.
%
\end{enumerate}
%
%
%
%

\end{enumerate}
\end{center}
}\end{minipage}
}
\end{center}

Putting the tie-breaking rule of $M_{Add}$ aside for a moment, let us first provide some intuition on why $A_{Th}$ works.
First notice that, for $\alpha$ from $n-n_i$ to $0$, $t_\alpha$ increases and $t'_\alpha$ decreases. Thus there is a unique value for $\alpha$ such that
$t_{\alpha+1}< t'_{\alpha+1}$ and $t_\alpha \geq t'_\alpha$. This will be where algorithm $A_{Th}$ stops.
The final value of $\alpha$ represents the largest number of units of all the other players that can appear before player $i$'s $j$-th unit, so that the latter can still be bought by $M_{Add}$. That is, at the end of the algorithm,  $\alpha+j$ represents the largest rank of $i$'s $j$-th unit so that it can be bought
(the first $j-1$ units of $i$ will always appear before his own $j$-th unit).

During the algorithm, for any $\alpha$, $t'_\alpha$ is the smallest cost player $i$ can announce so that his $j$-th unit appears after the $\alpha$-th unit of all the other players
in $M_{Add}$,
and $t_{\alpha}$ is the largest cost player $i$ can announce so that, when ranked $\alpha+j$, his $j$-th unit will satisfy the condition in Step \ref{enu:b} of $M_{Add}$.
When $t_\alpha< t'_\alpha$, there is no way for $i$'s $j$-th unit to be bought by $M_{Add}$ at rank $\alpha+j$.
When $t'_\alpha\leq t_\alpha\leq t'_{\alpha+1}$, by announcing anything in between $t'_\alpha$ and $t'_{\alpha+1}$ player $i$ has his $j$-th unit appearing after the others' $\alpha$-th unit but before their $\alpha+1$st unit,
and $t_\alpha$ is the largest cost so that his $j$-th unit will be bought.
Finally, when $t_\alpha > t'_{\alpha+1}$, by announcing $t_\alpha$ player $i$ will not have his $j$-th unit ranked $\alpha+j$: he must announce a smaller cost, and $t'_{\alpha+1}$ is the largest he can announce.

We formalize this intuition in the claim below.


\begin{claim}
The algorithm $A_{Th}$ computes the correct threshold $\theta_{ij}$.
\end{claim}
\begin{proof}
Since $t_0\geq 0 = t'_0$, the algorithm will always end in Steps \ref{enu:3b} or \ref{enu:3c} for some $\alpha$, and $\theta_{ij}$ will be set to some value.
Below we first show that if player $i$ bids $c'_i> \theta_{ij}$ then the mechanism will not pick up pair $(i, j)$. Indeed, if the algorithm stops in Step \ref{enu:3b}, then $\theta_{ij} = t_\alpha$ and $c'_i > t_\alpha\geq t'_\alpha$. Accordingly,
$$\frac{c'_i}{v_{ij}} > \frac{t'_\alpha}{v_{ij}} = \frac{c_{i'_\alpha}}{v_{i'_\alpha j'_\alpha}},$$
and pair $(i, j)$ is ranked after pair $(i'_\alpha, j'_\alpha)$ in the mechanism.
Therefore Inequality \ref{equ:2} is violated for pair $(i, j)$, because
$$\frac{c'_i}{v_{ij}} > \frac{t_\alpha}{v_{ij}} = \frac{B}{\sum_{\ell\leq j}v_{i\ell} + \sum_{\ell\leq \alpha}v_{i'_\ell j'_\ell}}.$$
Notice that pair $(i,j)$ may be ranked after pair $(i'_{\alpha+1}, j'_{\alpha+1})$ and so on, but then Inequality \ref{equ:2} remains violated,
since its right-hand side will only become smaller, with more terms added to the denominator.

If the algorithm instead stops in Step \ref{enu:3c}, then $c'_i > \theta_{ij} = t'_{\alpha+1}$, which implies
$$\frac{c'_i}{v_{ij}} > \frac{t'_{\alpha+1}}{v_{ij}} = \frac{c_{i'_{\alpha+1}}}{v_{i'_{\alpha+1} j'_{\alpha+1}}}.$$
Accordingly,
pair $(i, j)$ is ranked after pair $(i'_{\alpha+1}, j'_{\alpha+1})$. But the algorithm did not stop at $\alpha+1$, which means $t_{\alpha+1}< t'_{\alpha+1}$. Again Inequality \ref{equ:2} is violated for pair $(i,j)$, since
$$\frac{c'_i}{v_{ij}} > \frac{t'_{\alpha+1}}{v_{ij}} > \frac{t_{\alpha+1}}{v_{ij}} = \frac{B}{\sum_{\ell\leq j}v_{i\ell} + \sum_{\ell\leq \alpha+1}v_{i'_\ell j'_\ell}}.$$
Notice that pair $(i,j)$ may be ranked even further down,
but then Inequality \ref{equ:2} remains violated,
since its right-hand side will only become smaller.

In sum, pair $(i,j)$ will not be picked up by the mechanism for any $c'_i>\theta_{ij}$.
Next, we show if player $i$ bids $c'_i<\theta_{ij}$ then the mechanism will pick up pair $(i, j)$.
To do so, notice that no matter whether the algorithm stops in Step \ref{enu:3b} or \ref{enu:3c}, we have
$$c'_i < \theta_{ij} = \min\{t'_{\alpha+1}, t_\alpha\}.$$
Accordingly,
$$\frac{c'_i}{v_{ij}} < \frac{t'_{\alpha+1}}{v_{ij}} = \frac{c_{i'_{\alpha+1}}}{v_{i'_{\alpha+1} j'_{\alpha+1}}}$$
and
$$\frac{c'_i}{v_{ij}} < \frac{t_\alpha}{v_{ij}} = \frac{B}{\sum_{\ell\leq j}v_{i\ell} + \sum_{\ell\leq \alpha}v_{i'_\ell j'_\ell}}.$$
Thus $(i, j)$ is ranked before $(i'_{\alpha+1}, j'_{\alpha+1})$, and Inequality \ref{equ:2} is satisfied for $(i, j)$.
%
Again, $(i, j)$ may be ranked even earlier on, but then the right-hand side of Inequality \ref{equ:2} will only become larger, with some terms taken out from the denominator.
Thus $(i, j)$ will be picked up by the mechanism for any $c'_i< \theta_{ij}$.

Putting everything together, the claim holds. 
\end{proof}


The algorithm $A_{Th}$ clearly runs in polynomial time, thus the mechanism $M_{Add}$ also runs in polynomial time and Lemma \ref{lem:poly} holds. 
\end{proof}

\begin{lemma}\label{lem:budget}
Mechanism $M_{Add}$ is budget-feasible.
\end{lemma}
\begin{proof}
We shall prove an upper bound for the total payment made by the sub-mechanism in Step \ref{enu:1} of~$M_{Add}$, namely,
\begin{equation*}
\sum_{i\leq m}\sum_{j\leq a_i} \theta_{ij} \leq (1+\ln n)B.
\end{equation*}
To do so, recall that the mechanism picks up the first $k$ pairs in the ordered list according to the value-rates~$r_{ij}$'s, denoted by $(i_1, j_1)$ through $(i_k, j_k)$. (By definition $k = \sum_{i\leq m} a_i$.) 

Re-order these $k$ pairs according to $v_{ij}$'s decreasingly, with ties broken lexicographically. We denote by $(\hat{i}_s, \hat{j}_s)$
the $s$-th pair in this new ordering.
We have
\begin{equation}\label{equ:4}
v_{\hat{i}_1 \hat{j}_1} \geq v_{\hat{i}_2 \hat{j}_2}\geq \cdots \geq v_{\hat{i}_k \hat{j}_k}
\end{equation}
and
\begin{equation}\label{equ:5}
\sum_{\ell \leq k} v_{i_\ell j_\ell} = \sum_{\ell\leq k} v_{\hat{i}_\ell \hat{j}_\ell}.
\end{equation}
Below we show
$\theta_{\hat{i}_s \hat{j}_s} \leq \frac{B}{s}$ for any $s\leq k$.

Assume for the sake of contradiction that there exists $s\leq k$ such that
$\theta_{\hat{i}_{s} \hat{j}_{s}} > \frac{B}{s}$.
Consider the pair~$(i_k, j_k)$ in the mechanism's ordering. We have
$$\frac{c_{i_k}}{v_{i_k j_k}} \leq \frac{B}{\sum_{\ell\leq k} v_{i_\ell j_\ell}} = \frac{B}{\sum_{\ell \leq k}v_{\hat{i}_\ell \hat{j}_\ell}} \leq \frac{B}{\sum_{\ell\leq s}v_{\hat{i}_\ell \hat{j}_\ell}} \leq \frac{B}{s \cdot v_{\hat{i}_s \hat{j}_s}}< \frac{\theta_{\hat{i}_{s} \hat{j}_{s}}}{v_{\hat{i}_s \hat{j}_s}},$$
where the first inequality is by the construction of the mechanism, the equality is by Equation \ref{equ:5}, the second inequality is because $s\leq k$, the third inequality is by Equation \ref{equ:4}, and the last inequality is because $\theta_{\hat{i}_{s} \hat{j}_{s}} > \frac{B}{s}$. Therefore when player $\hat{i}_s$ bids some $c'_{\hat{i}_s} \in (\frac{B}{s}, \theta_{\hat{i}_{s} \hat{j}_{s}})$, we have
\begin{equation*}
\frac{c'_{\hat{i}_s}}{v_{\hat{i}_s \hat{j}_s}} > \frac{B}{s \cdot v_{\hat{i}_s \hat{j}_s}} \geq \frac{c_{i_k}}{v_{i_k j_k}}.
\end{equation*}
Accordingly, for any $\ell<s$ such that $\hat{i}_\ell \neq \hat{i}_s$, the pair $(\hat{i}_\ell, \hat{j}_\ell)$ is ranked before $(\hat{i}_s, \hat{j}_s)$ by the mechanism under player $\hat{i}_s$'s new bid $c'_{\hat{i}_s}$, because
$\frac{c_{\hat{i}_\ell}}{v_{\hat{i}_\ell \hat{j}_\ell}} \leq \frac{c_{i_k}}{v_{i_k j_k}}$.

Moreover, for any $\ell<s$ such that $\hat{i}_\ell = \hat{i}_s$, the pair $(\hat{i}_\ell, \hat{j}_\ell)$ is also ranked before~$(\hat{i}_s, \hat{j}_s)$ by the mechanism under player $\hat{i}_s$'s new bid, because $v_{\hat{i}_\ell \hat{j}_\ell} \geq v_{\hat{i}_s \hat{j}_s}$ and the two pairs have the same cost.
(When~$v_{\hat{i}_\ell \hat{j}_\ell} = v_{\hat{i}_s \hat{j}_s}$, it must be $\hat{j}_\ell < \hat{j}_s$ due to the lexicographic tie-breaking rule, and thus $(\hat{i}_\ell, \hat{j}_\ell)$ is still ranked before $(\hat{i}_s, \hat{j}_s)$ by the mechanism under player $\hat{i}_s$'s new bid, due to the same tie-breaking rule.)

Accordingly, when player $\hat{i}_s$ bids $c'_{\hat{i}_s}$, all $s-1$ pairs $(\hat{i}_1, \hat{j}_1), \dots, (\hat{i}_{s-1}, \hat{j}_{s-1})$ are ranked before~$(\hat{i}_s, \hat{j}_s)$ by the mechanism, and the total value of all pairs before or equal to $(\hat{i}_s, \hat{j}_s)$ is at least $\sum_{\ell\leq s} v_{\hat{i}_\ell \hat{j}_\ell}$. Because
$$\frac{c'_{\hat{i}_s}}{v_{\hat{i}_s \hat{j}_s}} > \frac{B}{s \cdot v_{\hat{i}_s \hat{j}_s}} \geq \frac{B}{\sum_{\ell\leq s}v_{\hat{i}_\ell \hat{j}_\ell}},$$
 the pair $(\hat{i}_s, \hat{j}_s)$ is not picked up by the mechanism under $c'_{\hat{i}_s}$, contradicting the fact that $c'_{\hat{i}_s} < \theta_{\hat{i}_{s} \hat{j}_{s}}$.

Therefore we have
$$\theta_{\hat{i}_s \hat{j}_s} \leq \frac{B}{s} \  \mbox{ for any } s\leq k,$$
which implies
$$\sum_{i\leq m}\sum_{j\leq a_i} \theta_{ij} = \sum_{s \leq k} \theta_{\hat{i}_s \hat{j}_s} \leq \sum_{s \leq k} \frac{B}{s} \leq B \sum_{s \leq n} \frac{1}{s} \leq (1+\ln n)B,$$
as we wanted to show. 

Since the sub-mechanism outputting $e_{i^*}$ pays $B$ to player $i^*$ and 0 to others, the expected payment of mechanism $M_{Add}$ is at most
$$\frac{1}{2(1+\ln n)} \cdot (1+\ln n) B + \frac{1}{2} \cdot B = B,$$
and Lemma \ref{lem:budget} holds. 
\end{proof}

\begin{lemma}\label{lem:approx}
Mechanism $M_{Add}$ is a $4(1+\ln n)$-approximation.
\end{lemma}
\begin{proof}
Recall that $A^*$ is an optimal allocation and $i^*$ is a player whose first unit has the highest value among all units of all players.
We shall show that $V(A) + V(e_{i*})$ is a 2-approximation of~$V(A^*)$, where $A$ is the output of Step \ref{enu:1}.

To do so, notice that once the cost profile $(c_1, \dots, c_n)$ is given, without strategic considerations, the optimization problem in our setting can be reduced to a 0-1 knapsack problem with $n$ items and budget~$B$. In particular,
for each player $i\in [m]$ and each $j\in [n_i]$, there is an item~$(i,j)$ with value $v_{ij}$ and cost $c_{ij} = c_i$. For any subset $S \subseteq \{(i,j): i\in [m], j\in [n_i]\}$, its value is $V(S) = \sum_{(i,j)\in S}v_{ij}$,
and its cost is $\sum_{(i,j)\in S}c_{ij} = \sum_{(i,j)\in S} c_i$. The allocation~$A^*$ naturally corresponds to an optimal set $S^*$ in the 0-1 knapsack problem, and the pair $(i^*, 1)$ is the item with the highest value.

For 0-1 knapsack, it is well known (see, e.g., \cite{Martello:1990:KPA:98124}) that the greedy algorithm that uses value-rate sorting and exhausts the budget
gives constant approximation. In particular, letting $(i_1, j_1), \dots, (i_n, j_n)$ be the ordered list with $r_{ij}$'s decreasing and ties broken lexicographically, and letting $\hat{k}$ be the largest number in $[n]$ satisfying $\sum_{\ell\leq \hat{k}} c_{i_\ell} \leq B$, we have
\begin{equation}\label{equ:6}
\sum_{\ell \leq \hat{k}} v_{i_\ell j_\ell} + v_{i^* 1} \geq \sum_{\ell \leq \hat{k}+1} v_{i_\ell j_\ell} \geq V(S^*) = V(A^*),
\end{equation}
where $v_{i_{\hat{k}+1} j_{\hat{k}+1}} = 0$ if $\hat{k} = n$.
The first inequality above is by the definition of $i^*$, and the second is because~$\sum_{\ell \leq \hat{k}+1} v_{i_\ell j_\ell}$ is greater than or equal to the optimal fractional solution.

Recall that the mechanism only picks up the first $k$ pairs and may not exhaust the budget.
To lower-bound the value generated by those $k$ pairs, notice that we have
$\frac{c_{i_\ell}}{v_{i_\ell j_\ell}} \leq \frac{c_{i_k}}{v_{i_k j_k}}$ for any $\ell \leq k$,
and thus
$$\frac{\sum_{\ell\leq k} c_{i_\ell}}{\sum_{\ell\leq k} v_{i_\ell j_\ell}} \leq \frac{c_{i_k}}{v_{i_k j_k}} \leq \frac{B}{\sum_{\ell\leq k} v_{i_\ell j_\ell}},$$
which implies $\sum_{\ell\leq k} c_{i_\ell} \leq B$. Accordingly, $k\leq \hat{k}$.

If $k=\hat{k}$, by Inequality \ref{equ:6} we have
\begin{equation}\label{equ:6-2}
V(A) + V(e_{i^*}) = \sum_{\ell \leq \hat{k}} v_{i_\ell j_\ell} + v_{i^* 1} \geq V(A^*).
\end{equation}
If $k< \hat{k}$, then
$$\frac{\sum_{k+1\leq \ell\leq \hat{k}} c_{i_\ell}}{\sum_{k+1 \leq \ell \leq \hat{k}} v_{i_\ell j_\ell}} \geq \frac{c_{i_{k+1}}}{v_{i_{k+1} j_{k+1}}} > \frac{B}{\sum_{\ell\leq k+1} v_{i_\ell j_\ell}},$$
where the first inequality is because $\frac{c_{i_\ell}}{v_{i_\ell j_\ell}} \geq \frac{c_{i_{k+1}}}{v_{i_{k+1} j_{k+1}}}$ for any $\ell \geq k+1$, and the second is by the construction of the mechanism.
Accordingly,
\begin{equation*}
\sum_{k+1 \leq \ell \leq \hat{k}} v_{i_\ell j_\ell} < \frac{\sum_{k+1\leq \ell\leq \hat{k}} c_{i_\ell}}{B} \cdot \sum_{\ell\leq k+1} v_{i_\ell j_\ell}\leq \sum_{\ell\leq k+1} v_{i_\ell j_\ell} \leq \sum_{\ell \leq k} v_{i_\ell j_\ell} + v_{i^* 1},
\end{equation*}
where the second inequality is because $\sum_{k+1\leq \ell\leq \hat{k}} c_{i_\ell} \leq \sum_{\ell \leq \hat{k}}c_{i_\ell} \leq B$,
and the last is because $v_{i_{k+1} j_{k+1}}\leq v_{i^* 1}$.
Thus
\begin{eqnarray}
2V(A) + 2V(e_{i^*}) &=& 2\sum_{\ell \leq k} v_{i_\ell j_\ell} + 2 v_{i^* 1} \geq \sum_{\ell \leq k} v_{i_\ell j_\ell} +  v_{i^* 1} + \sum_{k+1\leq \ell \leq \hat{k}} v_{i_\ell j_\ell} \nonumber \\
&=& \sum_{\ell \leq \hat{k}} v_{i_\ell j_\ell} +  v_{i^* 1} \geq V(A^*), \label{equ:7}
\end{eqnarray}
where the last inequality is by Inequality \ref{equ:6}.

Combining Inequalities \ref{equ:6-2} and \ref{equ:7},
we have that $V(A) + V(e_{i^*})$ is a $2$-approximation for $V(A^*)$. Thus the expected value of $M_{Add}$'s output is
$$\frac{1}{2(1+\ln n)}\cdot V(A) + \frac{1}{2} \cdot V(e_{i^*}) \geq \frac{V(A) + V(e_{i^*})}{2(1+\ln n)} \geq \frac{V(A^*)}{4(1+\ln n)}.$$
Since individually rationality and budget-feasibility have been shown by Lemmas \ref{lem:dst} and \ref{lem:budget},
mechanism $M_{Add}$ is a $4(1+\ln n)$-approximation and Lemma \ref{lem:approx} holds. 
\end{proof}

Theorem \ref{thm:additive} follows immediately from Lemmas \ref{lem:dst}-\ref{lem:approx}.

\section{Proofs of Theorems \ref{lem:one} and \ref{thm:sub}}\label{app:proof4}

\noindent
{\bf Theorem \ref{lem:one}} (restated){\bf .} {\em
Mechanism $M_{One}$ is universally truthful, individually rational, budget-feasible, and is a $(1+\ln n)$-approximation for $V(\lambda^{**} e_{i^{**}})$.
}

\begin{proof} To show that $M_{One}$ is universally truthful, we only need to show that the mechanism in Steps \ref{step:one 1}-\ref{step:one 4} is DST.
First of all, it is easy to see that the allocation is monotone. Indeed, for any player $i\neq i^{**}$, increasing~$i$'s cost can only cause $v_i$ to decrease.
Thus $i$ is still not the first in the list and sells 0 unit in the new allocation.
Decreasing his cost can only cause him to sell more units, since he sells~0 under $c_i$.

For player $i^{**}$, decreasing his cost to $c'< c_{i^{**}}$ can only cause $v_{i^{**}}$ to increase, thus he is still the first in the list, and the number of units he sells is $\min\{n_{i^{**}}, \lfloor\frac{B}{c'}\rfloor\} \geq \min\{n_{i^{**}}, \lfloor\frac{B}{c_{i^{**}}}\rfloor\}$.
On the other hand,
by increasing his cost to $c'> c_{i^{**}}$, he will either lose the first place and sell 0 unit, or still be the first but with the number of units $\min\{n_{i^{**}}, \lfloor\frac{B}{c'}\rfloor\} \leq \min\{n_{i^{**}}, \lfloor\frac{B}{c_{i^{**}}}\rfloor\}$. In sum, monotonicity holds.

\smallskip

Next, we show that for each $\ell \in [\lambda^{**}]$,
$\theta_\ell$ is the correct threshold for the $\ell$-th unit of player~$i^{**}$. We distinguish whether $\ell\leq k$ or not.

If $\ell\leq k$, then by bidding $c'_{i^{**}} > \theta_\ell = \frac{B}{k}$, we have $\frac{B}{c'_{i^{**}}}< k$, and thus $\lfloor\frac{B}{c'_{i^{**}}}\rfloor \leq k-1$. Accordingly,
$$V(\min\{n_{i^{**}}, \lfloor\frac{B}{c'_{i^{**}}}\rfloor\} \cdot e_i) \leq V(\min\{n_{i^{**}}, k-1\} \cdot e_i) = V(\min\{n_{i^{**}}, \lfloor\frac{B}{B/(k-1)}\rfloor\} \cdot e_i).$$
By the definition of $k$, player $i^{**}$ is not the first in the list by bidding $\frac{B}{k-1}$.
Thus by bidding $c'_{i^{**}}$ he is not the first either,
and does not sell his $\ell$-th unit.

By bidding $c'_{i^{**}}< \theta_\ell$, we have $\frac{B}{c'_{i^{**}}} \geq k$, and thus
$$V(\min\{n_{i^{**}}, \lfloor\frac{B}{c'_{i^{**}}}\rfloor\} \cdot e_i) \geq V(\min\{n_{i^{**}}, \lfloor\frac{B}{B/k}\rfloor\} \cdot e_i).$$
Since player $i^{**}$ is the first by bidding $\frac{B}{k}$, he is still the first by bidding $c'_{i^{**}}$,
and the number of units he sells is $\min\{n_{i^{**}}, \lfloor\frac{B}{c'_{i^{**}}}\rfloor\} \geq \min\{n_{i^{**}}, \lfloor\frac{B}{B/k}\rfloor\} = k \geq \ell$,
where the equality is because $k\leq \lambda^{**} \leq n_{i^{**}}$.
That is, by bidding $c'_{i^{**}}$ he still sells his $\ell$-th unit.
Therefore $\theta_\ell$ is the correct threshold.

If $k+1\leq \ell \leq \lambda^{**}$, then by bidding $c'_{i^{**}} > \theta_\ell = \frac{B}{\ell}$ player $i^{**}$ will not sell his $\ell$-th unit even if he remains to be the first in the list, since $\lfloor\frac{B}{c'_{i^{**}}}\rfloor \leq \ell-1$. By bidding $c'_{i^{**}} < \theta_\ell$, we have $V(\min\{n_{i^{**}}, \lfloor\frac{B}{c'_{i^{**}}}\rfloor\} \cdot e_i) \geq V(\min\{n_{i^{**}}, \lfloor\frac{B}{B/\ell}\rfloor\} \cdot e_i)$. Again by the definition of $k$, by bidding $\frac{B}{\ell}$ player $i$ is the first in the list, and thus by bidding $c'_{i^{**}}$ he is still the first. The number of units he sells is $\min\{n_{i^{**}}, \lfloor\frac{B}{c'_{i^{**}}}\rfloor\} \geq \min\{n_{i^{**}}, \lfloor\frac{B}{B/\ell}\rfloor\} = \ell$. That is, by bidding $c'_{i^{**}}$ he still sells his $\ell$-th unit.

In sum, the $\theta_\ell$'s are the correct thresholds, the mechanism in Steps \ref{step:one 1}-\ref{step:one 4} is DST, and mechanism $M_{One}$ is universally truthful.

\smallskip

Individual rationality of $M_{One}$ follows from the fact that
the mechanism in Steps \ref{step:one 1}-\ref{step:one 4} is individually rational.
Indeed, for each $\ell\leq \lambda^{**}$ we have
$$\theta_\ell = \min\{\frac{B}{k}, \frac{B}{\ell}\} \geq \frac{B}{\lambda^{**}} \geq \frac{B}{\lfloor\frac{B}{c_{i^{**}}}\rfloor} \geq c_{i^{**}}.$$
Thus $\sum_{\ell \leq \lambda^{**}}\theta_\ell \geq \lambda^{**} c_{i^{**}}$, player $i^{**}$ has non-negative utility, and $M_{One}$ is individually rational.

Furthermore, because for each $\ell\leq \lambda^{**}$ we have $\theta_\ell \leq \frac{B}{\ell}$, the total payment made in Step \ref{step:one 4} is
$$\sum_{\ell \leq \lambda^{**}}\theta_\ell \leq \sum_{\ell \leq \lambda^{**}} \frac{B}{\ell} \leq \sum_{\ell \leq n} \frac{B}{\ell} \leq (1+\ln n)B.$$
Since this payment is made with probability $\frac{1}{1+\ln n}$, mechanism $M_{One}$ is budget-feasible in expectation.

Finally, under the true cost profile, mechanism $M_{One}$ outputs $\lambda^{**} e_{i^{**}}$ with probability $\frac{1}{1+\ln n}$, and thus is a $(1+\ln n)$-approximation for $V(\lambda^{**} e_{i^{**}})$. In sum, Theorem \ref{lem:one} holds. 
\end{proof}

\medskip

\noindent
{\bf Theorem \ref{thm:sub}} (restated) {\bf .} {\em
Mechanism $M_{Sub}$ runs in polynomial time, is universally truthful and individually rational, and is an $O(\frac{(\log n)^2}{\log \log n})$-approximation for procurement games with sub-additive valuations.
}

\smallskip

In order to prove Theorem \ref{thm:sub}, first notice that the mechanism $M_{Sub}$ clearly runs in polynomial time.
The proof that mechanism $M_{Rand}$ is universally truthful, individually rational, and budget-feasible is almost the same as in \cite{Bei12}, and thus we omit it here.
By Theorem \ref{lem:one}, mechanism $M_{One}$ also satisfies all of those properties. Thus mechanism $M_{Sub}$ is universally truthful, individually rational, and budget-feasible.
It remains to analyze the approximation ratio of $M_{Sub}$, 
%
and we proceed by first proving the following three lemmas. 
The framework of the analysis follows from \cite{Bei12}, but many new ideas are needed. In particular the proof of Lemma \ref{lem:log} requires novel ways of dealing with multi-unit allocations.

\begin{lemma}\label{lem:max}
For any input to $A_{Max}$, letting $\hat{A}$ be the optimal allocation under the same input, 
we have $V(S_{Max}) \geq \frac{V(\hat{A})}{8}$. 
\end{lemma}
%
%

\begin{proof}
Denoting $\hat{A}$ by $(\hat{a}_1,\dots, \hat{a}_m)$, we have $\hat{A} = \bigvee_{i\in [m]} \hat{a}_i e_i$, and sub-additivity implies
$$V(\hat{A}) \leq \sum_{i\in [m]} V(\hat{a}_i e_i).$$
Since $\sum_i \hat{a}_i c_i \leq B$, for each player $i$ we have $\hat{a}_i c_i \leq B$, implying
$$\hat{a}_i\leq \min\{n_i, \lfloor\frac{B}{c_i}\rfloor\} = n'_i,$$
and further implying $V(\hat{a}_i e_i) \leq V(n'_i e_i) \leq V(n'_{i^{**}} e_{i^{**}})$. Accordingly,
$$V(\hat{A})\leq m V(n'_{i^{**}} e_{i^{**}}) = m v^*.$$
Since on the other hand we have $V(\hat{A})\geq V(n'_{i^{**}} e_{i^{**}}) = v^*$, there exists $v\in \cV$ such that $\frac{V(\hat{A})}{2}\leq v \leq V(\hat{A})$.

Fix such a $v$ and let $S = (s_1,\dots, s_m)$ be the allocation returned by the demand oracle for $v$. We have
%
$$V(S) - \frac{v}{2B}\cdot \sum_{i\in [m]} s_i c_i \geq V(\hat{A}) - \frac{v}{2B}\cdot \sum_{i\in [m]} \hat{a}_i c_i \geq v - \frac{v}{2B}\cdot B = \frac{v}{2},$$
thus $V(S)\geq \frac{v}{2}$ and algorithm $A_{Max}$ will not output $A_{\bot}$. If $\sum_i s_i c_i\leq B$, then $S_v = S$ and
$$V(S_{Max}) \geq V(S_v) \geq \frac{v}{2} \geq \frac{V(\hat{A})}{4}\geq \frac{V(\hat{A})}{8},$$
as desired. 
%

\smallskip

Assume now $\sum_i s_i c_i > B$. 
By the construction of $S_v$ we have $\sum_{\ell\leq k}s_{i_\ell}c_{i_\ell} > \frac{B}{2}$.
Letting $S'_v = \bigvee_{k+1\leq \ell\leq m} s_{i_\ell} e_{i_\ell}$ , we have $S = S_v \vee S'_v$, and thus
$V(S) \leq V(S_v) + V(S'_v)$, which implies
\begin{equation}
V(S) - \frac{v}{2B}\cdot \sum_{i\in [m]} s_i c_i  \leq   V(S_v) - \frac{v}{2B}\cdot \sum_{\ell\leq k}s_{i_\ell}c_{i_\ell} + V(S'_v) - \frac{v}{2B}\cdot \sum_{k+1\leq\ell\leq m}s_{i_\ell}c_{i_\ell}.
\end{equation}
If $V(S_v) < \frac{v}{4}$, then 
\begin{eqnarray*}
V(S) - \frac{v}{2B}\cdot \sum_{i\in [m]} s_i c_i & < & \frac{v}{4} - \frac{v}{2B} \cdot \frac{B}{2} + V(S'_v) - \frac{v}{2B}\cdot \sum_{k+1\leq\ell\leq m}s_{i_\ell}c_{i_\ell} \\
&=& V(S'_v) - \frac{v}{2B}\cdot \sum_{k+1\leq\ell\leq m}s_{i_\ell}c_{i_\ell},
\end{eqnarray*}
contradicting the fact that $S$ is the optimal solution returned by the demand oracle.
Thus
$$V(S_v)\geq \frac{v}{4}\geq \frac{V(\hat{A})}{8},$$
implying $V(S_{Max}) \geq \frac{V(\hat{A})}{8}$. 

Thus Lemma \ref{lem:max} holds. 
\end{proof}

%
%

%

Next, consider the two sets of players $T$ and $T'$ in Step \ref{enu:rand-1} of $M_{Rand}$.
Let $\hat{A}_T$ and $\hat{A}_{T'}$ respectively be the optimal allocation among the budget-feasible ones that only take units from players in $T$ and $T'$,
and let $A^*$ be the optimal allocation whose social welfare $M_{Sub}$ aims to approximate. We have the following. 

\begin{lemma}\label{lem:TandT'}
With probability at least $1/4$,
\begin{equation}\label{equ:5-2}
V(\hat{A}_{T'})\geq V(\hat{A}_T) \geq \frac{V(A^*)}{8}.
\end{equation}
\end{lemma}

%
%

\begin{proof}
For any subset of players $C$, let $A^*_C$ be $A^*$ projected to $C$: letting $A^* = (a^*_1,\dots, a^*_n)$, $A^*_C =\bigvee_{i\in C} a^*_i e_i$.
If $C = \{i\}$, we write $A^*_i$ instead of $A^*_{\{i\}}$.
We show that there exists two disjoint player sets $C_1, C_2$ such that $C_1 \cup C_2 = [m]$,
$$V(A^*_{C_1}) \geq \frac{V(A^*)}{4}, \mbox{ and } V(A^*_{C_2}) \geq \frac{V(A^*)}{4}.$$

We start with $C_1 = \emptyset$ and $C_2 = [m]$, and move players to $C_1$ one by one in an arbitrary order, until $V(A^*_{C_1}) \geq \frac{V(A^*)}{4}$. Letting $i$ be the last player moved, we have $V(A^*_{C_1\setminus\{i\}}) < \frac{V(A^*)}{4}$. Since $C_1\setminus\{i\}$ and $C_2\cup\{i\}$ are two disjoint sets whose union is $[m]$, we have
$$A^* = A^*_{C_1\setminus\{i\}}\vee A^*_{C_2\cup\{i\}},$$
and sub-additivity implies $V(A^*) \leq V(A^*_{C_1\setminus\{i\}}) + V(A^*_{C_2\cup\{i\}})$.\footnote{Notice that in single-unit settings partitioning the players is the same as partitioning the units, and given the optimal allocation $S^*$ which is a subset of players, for any set $S\subseteq S^*$ we have $V(S^*) \leq V(S) + V(S^*\setminus S)$. But in our setting the partition has to be done in terms of players rather than units. Indeed, partitioning $A^* = (a^*_1,\dots, a^*_m)$ into two arbitrary allocations $A = (a_1,\dots, a_m)$ and $A' = (a'_1, \dots, a'_m)$ with $a_i+a'_i =a^*_i$ for each $i$ will not give us $A^* = A\vee A'$, since there may exist $i$ such that both $a_i$ and $a'_i$ are strictly less than $a^*_i$. Only when $A$ and $A'$ are $A^*$ projected to two disjoint player sets will one have $A^* = A\vee A'$ and $V(A^*) \leq V(A) + V(A')$.}
Accordingly $V(A^*_{C_2\cup\{i\}}) > \frac{3V(A^*)}{4}$.
Since $A^*$ is budget-feasible, the number of units $i$ sells in $A^*$ is at most $\min\{n_i, \lfloor\frac{B}{c_i}\rfloor\}$, and we have
$$V(A^*_i) \leq V(\min\{n_i, \lfloor\frac{B}{c_i}\rfloor\}\cdot e_i) \leq V(\lambda^{**} e_{i^{**}}) < \frac{V(A^*)}{2},$$
where the first inequality is because $V$ is monotone and the second is by the definition of $i^{**}$. Since~$i\notin C_2$, we have $A^*_{C_2\cup\{i\}} = A^*_{C_2} \vee A^*_i$, and again
sub-additivity implies
$V(A^*_{C_2\cup\{i\}}) \leq V(A^*_{C_2}) + V(A^*_i)$. Thus
$$V(A^*_{C_2}) \geq V(A^*_{C_2\cup\{i\}}) - V(A^*_i) > \frac{3V(A^*)}{4} - \frac{V(A^*)}{2} = \frac{V(A^*)}{4}.$$
Since $T\cap T' = \emptyset$ and $T\cup T' = [m]$, for each $C_i$ we have $A^*_{C_i} = A^*_{C_i\cap T}\vee A^*_{C_i\cap T'}$, and thus $V(A^*_{C_i}) \leq V(A^*_{C_i\cap T}) + V(A^*_{C_i\cap T'})$. Accordingly, either $V(A^*_{C_i\cap T})$ or $V(A^*_{C_i\cap T'})$ is greater than~$\frac{V(A^*)}{8}$. Since the players in $C_1$ and $C_2$ are partitioned to $T$ and $T'$ uniformly and independently, with probability at least $1/2$ there exists $i$ such that
\begin{equation}\label{equ:5-1}
V(A^*_{C_i \cap T})\geq \frac{V(A^*)}{8} \mbox{ and } V(A^*_{C_{3-i} \cap T'}) \geq \frac{V(A^*)}{8},
\end{equation}
namely, the more valuable parts of $C_1$ and $C_2$ end up at different sides.

Since both allocations $A^*_{C_i \cap T}$ and $A^*_{C_{3-i} \cap T'}$ in Equation \ref{equ:5-1} are budget-feasible, we have
$V(\hat{A}_T) \geq V(A^*_{C_i \cap T})$ and $V(\hat{A}_{T'}) \geq V(A^*_{C_{3-i} \cap T'})$. Thus with probability at least $1/2$, $V(\hat{A}_T) \geq \frac{V(A^*)}{8}$ and $V(\hat{A}_{T'}) \geq \frac{V(A^*)}{8}$. Because the role of $T$ and $T'$ can be switched, with probability $1/2$ we have $V(\hat{A}_{T'}) \geq V(\hat{A}_{T})$. Thus with probability at least $1/4$ we have $V(\hat{A}_{T'})\geq V(\hat{A}_T) \geq \frac{V(A^*)}{8}$, and Lemma \ref{lem:TandT'} holds. 
\end{proof}

Furthermore, recall that $v$ is the value computed in Step \ref{enu:rand-2} of Mechanism $M_{Rand}$ and
$(i^{**}, \lambda^{**})$ is the optimal single-item allocation. Letting $A$ be the outcome of $M_{Rand}$, we have the following.

\begin{lemma}\label{lem:log}
When Inequality \ref{equ:5-2} holds, $V(A) + V(\lambda^{**} e_{i^{**}}) \geq \frac{\log\log n}{64\log n} \cdot v$.
\end{lemma}

%

\begin{proof}

We shall partition $\hat{A}_{T'}$ into disjoint sets. But instead of partitioning according to the players as we have done in Lemma \ref{lem:TandT'}, this time we shall partition according to the units.
Let $t = |T'|$, $\hat{a}_i$ be the number of units each player $i\in T'$ sells in $\hat{A}_{T'}$, and $n' = \sum_{i\in T'} \hat{a}_i$. Without loss of generality, assume $T' = \{1, 2, \dots, t\}$ and $c_1\geq c_2\geq \cdots \geq c_t$.

Let $L$ be the ordered list of player-unit pairs
$$(1, 1),\dots, (1, \hat{a}_1), (2, 1), \dots, (2, \hat{a}_{2}), \dots, (t, 1), \dots, (t, \hat{a}_{t}),$$
 and denote by $(i_\ell, j_\ell)$ the $\ell$-th pair in $L$, with $\ell \in [n']$.
We recursively partition the pairs in $L$ into different groups as follows:
\begin{itemize}

\item
Let $\alpha_1$ be the largest integer such that $c_{i_1} \leq \frac{B}{\alpha_1}$. Put the first $\alpha'_1 = \min\{\alpha_1, n'\}$ pairs into group $Z_1$.

\item
Let $\beta_r = \alpha'_1 + \cdots + \alpha'_r$. If $\beta_r< n'$, then let $\alpha_{r+1}$ be the largest integer such that $c_{i_{\beta_r +1}} \leq \frac{B}{\alpha_{r+1}}$. Put the next $\alpha'_{r+1} = \min\{\alpha_{r+1}, n'-\beta_r\}$ pairs in group $Z_{r+1}$.

\end{itemize}
Let $x+1$ be the number of groups. For each $r\in [x+1]$, notice that $Z_r$ naturally correspond to an allocation where each player $i$'s number of units is the number of pairs of his in $Z_r$. Slightly abusing notation, we refer to this allocation as $Z_r$ as well, and use $V(Z_r)$ to denote its value.

If $x=0$, then there is only one group $Z_1 = L$. Thus $Z_1 = \hat{A}_{T'}$ and $V(Z_1) = V(\hat{A}_{T'})$.
In round $k = \alpha_1$ of mechanism $M_{Rand}$, we have $T_k = T'$ since $c_i \leq c_1 \leq \frac{B}{\alpha_1}$ for each $i\in T'$. The optimal budget-feasible allocation for $T_k$ with unit-cost $\frac{B}{\alpha_1}$ has value at least $V(Z_1)$,
because $|Z_1|\leq \alpha_1$, which makes $Z_1$ a budget-feasible allocation under unit-cost $\frac{B}{\alpha_1}$. By Lemma \ref{lem:max}, in this round we have
$$V(X) \geq \frac{V(Z_1)}{8} = \frac{V(\hat{A}_{T'})}{8} \geq \frac{\log\log n}{8\log n} \cdot v,$$
where the last inequality is because $v$ is the value of a budget-feasible allocation for players in $T$,
which implies $V(\hat{A}_T) \geq v$, and thus
$$V(\hat{A}_{T'})\geq v$$
 by Inequality \ref{equ:5-2}.
Thus the mechanism, which may terminate before or at round $\alpha_1$, will output an allocation $A$ such that $V(A) \geq \frac{\log\log n}{64 \log n}\cdot v$, and Lemma~\ref{lem:log} holds.

If $x>1$, notice that 
%
for any $1\leq r< x$ there is at most one player whose pairs appear in both $Z_r$ and $Z_{r+1}$: he is the last one picked up by $Z_r$ and the first by $Z_{r+1}$. Denote this player by $j_r$.\footnote{In principle it is possible that a player's units spread among several consecutive groups $Z_r, Z_{r+1}, Z_{r+2}, \dots$. In this case any group in the middle contains only pairs of this player, and $j_r = j_{r+1} = \dots$. This will not affect our analysis.}
We have
$$\hat{A}_{T'} = Z_1 \vee \hat{a}_{j_1} e_{j_1} \vee Z_2 \vee \hat{a}_{j_2} e_{j_2} \vee \cdots \vee \hat{a}_{j_x} e_{j_x} \vee Z_{x+1},$$
where $\hat{a}_{j_r} e_{j_r}$ is defined to be $A_\bot$ if there is no such a player $j_r$ between some $Z_r$ and $Z_{r+1}$. This is because, for any player in $T'$, either all his units taken by $\hat{A}_{T'}$ appear in some $Z_r$, or he is player $j_r$ for some $r$ and all his units appear in $\hat{a}_{j_r} e_{j_r}$. By sub-additivity,
\begin{equation}\label{equ:8-2}
V(\hat{A}_{T'}) \leq \sum_{r\in [x+1]} V(Z_r) + \sum_{r\in [x]} V(\hat{a}_{j_r} e_{j_r}) \leq \sum_{r\in [x+1]} V(Z_r) + x V(\lambda^{**} e_{i^{**}}),
\end{equation}
where the second inequality is because $\hat{A}_{T'}$ is budget-feasible, and thus for each $i\in T'$ we have $\hat{a}_i c_i\leq B$, implying $\hat{a}_i \leq \min\{n_i, \lfloor\frac{B}{c_i}\rfloor\}$.

Letting $r^* \in \argmax_{r\in [x+1]} V(Z_r)$, by Inequality \ref{equ:8-2} we have
$$V(\hat{A}_{T'}) \leq x V(\lambda^{**} e_{i^{**}}) + (x+1) V(Z_{r^*}).$$
By a similar argument as in \cite{Bei12}, we have $n'\geq \left(\frac{x}{2}\right)^x$, which implies $n \geq \left(\frac{x}{2}\right)^x$, and thus $x \leq \frac{2\log n}{\log \log n}$.
Accordingly,
$$V(\lambda^{**} e_{i^{**}}) + V(Z_{r^*}) \geq \frac{V(\hat{A}_{T'})}{x+1} \geq \frac{V(\hat{A}_{T'})}{2x} \geq \frac{\log \log n}{4\log n} \cdot V(\hat{A}_{T'}).$$
If $V(\lambda^{**} e_{i^{**}}) \geq \frac{\log \log n}{8\log n} \cdot V(\hat{A}_{T'})$, then Lemma \ref{lem:log} holds immediately, again because $V(\hat{A}_{T'})\geq v$. Otherwise, we have
$V(Z_{r^*}) \geq \frac{\log \log n}{8\log n} \cdot V(\hat{A}_{T'})$.
In round $k = \alpha_{r^*}$ of mechanism $M_{Rand}$, $T_k$ includes all players whose pairs appear in $Z_{r^*}$, and thus the optimal budget-feasible allocation for $T_k$ with unit-cost $\frac{B}{\alpha_{r^*}}$ is at least $V(Z_{r^*})$, because $|Z_{r^*}|\leq \alpha_{r^*}$, which makes $Z_{r^*}$ a budget-feasible allocation for $T_k$ with unit-cost $\frac{B}{\alpha_{r^*}}$.
By Lemma \ref{lem:max}, the allocation $X$ in this round satisfies
$$V(X) \geq \frac{V(Z_{r^*})}{8} \geq \frac{\log \log n}{64\log n} \cdot V(\hat{A}_{T'}) \geq \frac{\log \log n}{64\log n} \cdot v.$$
Thus the mechanism, which may terminate before or at round $\alpha_{r^*}$, will output an allocation $A$ such that $V(A) \geq \frac{\log\log n}{64 \log n}\cdot v$, and Lemma~\ref{lem:log} holds. 
\end{proof}

At this point, we are ready to prove Theorem \ref{thm:sub}.

\begin{proof}[of Theorem \ref{thm:sub}]
By Lemma \ref{lem:max}, the value $v$ satisfies
$v\geq \frac{V(\hat{A}_T)}{8}$.
When Inequality \ref{equ:5-2} holds, we have $V(\hat{A}_T)\geq \frac{V(A^*)}{8}$, and thus  
\begin{equation}\label{equ:7-1}
v \geq \frac{V(A^*)}{64}.
\end{equation}

By Lemma \ref{lem:TandT'}, Inequality \ref{equ:5-2} holds with probability at least $1/4$. Thus by Theorem \ref{lem:one}, Lemma \ref{lem:log} and Inequality \ref{equ:7-1}, the expected value generated by $M_{Sub}$ under the true cost profile is
at least
\begin{eqnarray*}
& & \frac{1}{4} \cdot \left(\frac{V(A)}{2} + \frac{V(\lambda^{**} e_{i^{**}})}{2(1+\ln n)}\right) \geq \frac{V(A) + V(\lambda^{**} e_{i^{**}})}{8(1+\ln n)}
\geq \frac{1}{8(1+\ln n)}\cdot \frac{\log \log n}{64\log n} \cdot v \\
&\geq& \frac{1}{8(1+\ln n)}\cdot \frac{\log \log n}{64\log n} \cdot \frac{V(A^*)}{64} = \frac{V(A^*)}{O(\frac{(\log n)^2}{\log\log n})}.
\end{eqnarray*}
Thus mechanism $M_{Sub}$ is an $O(\frac{(\log n)^2}{\log\log n})$-approximation and Theorem \ref{thm:sub} holds. 
\end{proof}

\end{document}